\documentclass[seceq]{ptptex}
\usepackage{graphicx} 
\usepackage{latexsym}
\notypesetlogo
\begin{document}
 \begin{flushright}
   RUP 11-3 \\
\end{flushright}
\vspace{10mm} 

\def\proj{{\bf P}}  
\def\slsh#1{{#1}{\kern-6pt}/{\kern1pt}}  
\begin{center}
\Large{  NLL Order Contributions for Exclusive Processes \\ in Jet-Calculus Scheme  }

\vspace{10mm}

\large{Hidekazu {\sc Tanaka}\\

\vspace{3mm}

Department of Physics, Rikkyo University, Tokyo 171-8501, Japan \\
     }


\vspace{15mm}

  {\large  \bf Abstract}
  \end{center}
        
\vspace{10mm}

We investigate the next-to-leading logarithmic (NLL) order contributions of the quantum chromodynamics  (QCD) for exclusive processes evaluated by  Monte Carlo methods.
Ambiguities of the Monte Carlo calculation based on the leading-logarithmic (LL) order approximations are pointed out.  
To remove these ambiguities, we take into account the NLL order terms.
 In a model presented in this paper, interference contributions due to the NLL order terms are included for the generation of the transverse momenta in initial-state parton radiations. Furthermore, a kinematical constraint due to parton radiation, which is also a part of the NLL order contributions, is taken into account.  This method guarantees a proper phase space boundary for hard scattering cross sections as well as parton radiations.
  As an example, cross sections for lepton pair productions mediated by a virtual photon in hadron-hadron collisions are calculated, using the jet-calculus scheme for flavor nonsinglet quarks.    

\newpage


\section{Introduction}

Monte Carlo methods with parton shower models for initial-state parton radiations are powerful tools for the  evaluation of exclusive processes for hadron-hadron scatterings.   
The parton shower models have been constructed  on the basis of perturbative  quantum chromodynamics (QCD). \cite{rf:1}
  
In actual calculations, the leading-logarithmic (LL) order of QCD is insufficient to evaluate the hadron-hadron scattering processes due to large parameter dependence, particularly the choice of a factorization scale parameter.  Thus, the next-to-leading logarithmic (NLL) order contributions\cite{rf:2}\footnote{In this paper, the NLL order contributions correspond to the corrections that appear from $O(\alpha_s^2)$ terms of splitting functions in parton evolutions,  as well as  boundary conditions due to momentum conservations at the collinear parton radiations.  Furthermore, $O(\alpha_s^2\log K^2)$ terms of the decay matrix elements (three-body decay functions) for branching vertices are also included. Here, $-K^2$ is a virtuality of an initial-state parton momentum.} should be taken into account. The next-to-leading order (NLO)   calculation ($O(\alpha_s)$ contributions for hard processes) is also necessary to remove theoretical ambiguities due to the factorization procedure as well as the choice of the factorization scale. \cite{rf:2} Here, $\alpha_s$ is the coupling constant of QCD.

 Conventionally, the scaling violation of the parton distributions is  calculated by solving the renormalization group equations in moments. Then, these solutions are numerically inverted to yield momentum fractions of partons.    
    Alternatively, we can use parton shower models in order to evaluate the scaling violation of the parton distributions.  One such algorithm has been proposed in Refs. \citen{rf:3} and \citen{rf:4}.  In this model, the scaling violation of the parton distributions is generated using only information from the splitting functions of the parton branching vertices and input distributions at a given energy.  It has been found that the method reproduces the scaling violation of the parton distributions up to their normalizations at the NLL order of QCD.  

  In order to allow the application of parton shower models to realistic processes, the matching problem should be solved, that is, double counting between the hard scattering and the parton showers must be avoided. So far, various methods have been proposed in order to solve this problem.$^{5)-9)}$  In particular, a systematic investigation of the factorization procedure at the NLO in Monte Carlo calculation has been performed, as described in Ref. \citen{rf:5}.  
In a recent work, a modified subtraction scheme at the NLO has been proposed.\cite{rf:9}.  They derived  hard scattering cross sections with the new scheme for the Drell-Yan process as well as for the deep inelastic scattering process.  Another method has been proposed in Ref. \citen{rf:6}, in which the collinear terms are subtracted from matrix elements before integration over the phase space.   However, in these works, the initial-state parton evolutions are based on the LL order of QCD.   
 
 In the conventional $\overline{\rm MS}$ scheme,\cite{rf:2}  it has been pointed out that matching between the hard scattering cross section and the initial-state parton radiation is broken in the calculation of exclusive processes.\cite{rf:7}
 In order to implement a factorization scheme that is appropriate for the evaluation of exclusive processes  to the accuracy of the NLL order of QCD, a kinematical constraint due to parton radiation is taken into account when considering the subtraction terms, which is called the $\overline{\rm MS}'$ scheme. The momenta of partons in the scattering processes are conserved in this method.  As an example, cross sections for the Drell-Yan lepton-pair production mediated by a virtual photon in hadron-hadron collisions have been calculated.\cite{rf:7} 

 However, in the collinear region after the collinear singularity is subtracted,  negative contributions remain for the cross section obtained with both the conventional $\overline{\rm MS}$ scheme and the $\overline{\rm MS}'$ scheme.\cite{rf:8}  Such contributions cannot be ignored at the NLL order accuracy. Event generation employed in the Monte Carlo methods with a negative probability may not be appropriate, since the strong cancellation between the negative contributions from the hard scattering cross section and the positive contributions from parton showers may give unstable results.

To solve this problem, the jet calculus (${\rm JC}$) scheme has been implemented in Refs. \citen{rf:8} and \citen{rf:10}.  The ${\rm JC}$ algorithm has been proposed by Konishi et al. in order to evaluate multiparton productions due to perturbative QCD. \cite{rf:11} This algorithm had been applied to  jet  evolutions as well as initial-state parton radiations.  Furthermore, for the parton evolution in which the virtualites of partons are strongly ordered, additional phase space factors, compared with the $\overline{\rm MS}$ scheme, are subtracted from the hard scattering cross sections. \cite{rf:12}

In this paper, the NLL order contributions in initial-state parton radiations as well as the NLO corrections in the hard scattering cross section  for exclusive processes using a Monte Carlo method are investigated in a consistent manner.  As an example, we calculate Drell-Yan lepton-pair production in proton-antiproton scattering with the ${\rm JC}'$ scheme, where a kinematical constraint due to parton radiation for the hard scattering process is also taken into account. In this algorithm, most of the collinear contributions to the hard scattering cross section at the NLO are subtracted. \cite{rf:8}  In order to simplify its explanation, we concentrate on the quark $(q)$--antiquark $({\bar q})$ annihilation process, but we include following ingredients for the evaluation of the physical cross sections:

\vspace{5mm}

\begin{description}
\item{(1)} Parton showers for initial-state radiations are generated at the NLL order accuracy including the $O(\alpha_s^2)$ terms for the splitting function,\cite{rf:13} where the $O(\alpha_s^2\log K^2)$ contributions are also taken into account for each branching step. The scaling violation is reproduced by evolution of the initial-state parton radiations at the NLL order accuracy up to the normalization of the distribution function, as well as the transverse momentum of partons.\cite{rf:4,rf:10}
\item{(2)} Explicit expressions of the three-body decay functions ($O(\alpha_s^2\log K^2)$ terms of the decay matrix elements in the parton branching vertices) are included, where interference contributions of two gluon radiations are also taken into account in the exact form.\cite{rf:14} 
\item{(3)} Kinematical constraints due to the momentum conservation of parton radiations are taken into account for both factorization of mass singularity for  the hard scattering cross section and the parton generations.\cite{rf:7,rf:8}
\end{description}

\vspace{5mm}
Since the above contributions are related to each other at the NLL order accuracy, one should  treat these terms consistently. Otherwise, the factorization scheme invariance for the physical cross sections may be broken. 
 
In $\S$2, we explain our algorithm.  
 Some numerical results are shown in $\S$3. Section 4 contains a summary and some comments.   The explicit expressions of the formula implemented in our algorithm are presented in Appendices.

\section{Basic formula for Monte Carlo algorithm}

Here, we consider the Drell-Yan lepton-pair production in quark $(q)$--antiquark $({\bar q})$ annihilation in a proton-antiproton scattering
\begin{eqnarray}
 q(p_q)+{\bar q}(p_{\bar q}) \rightarrow \gamma^*(q)+g(p_g) \rightarrow \l^-(p_-)+l^+(p_+)+g(p_g), 
 \end{eqnarray}
mediated by a photon $\gamma^*$ with the virtuality $(p_- + p_+)^2=q^2=Q^2$, where a gluon $(g)$ is radiated in the final state. Here, $p_i~(i=q,{\bar q},g)$ and $p_{\pm}$ denote the momenta of the corresponding particles.\footnote{The Mandelstam variables are defined by ${\hat s}=(p_q+p_{\bar q})^2,~~{\hat t}=(p_q-p_g)^2$, and ${\hat u}=(p_{\bar q}-p_g)^2 $, which satisfy ${\hat s}+{\hat t}+{\hat u}=Q^2$ for massless partons.}

\subsection{Subtraction scheme}

In order to obtain a finite cross section for the process $q{\bar q}\rightarrow \gamma^*g$, we subtract collinear contributions due to branching processes $q \rightarrow qg$ and ${\bar q} \rightarrow {\bar q}g$ from the hard scattering cross section.  Although these contributions  are compensated by initial-state parton radiations, the remnant of the subtracted cross section in the NLO cannot be ignored at the NLL order accuracy.  Here, we consider valence quarks (flavor nonsinglet sector) in order to avoid the quark-gluon mixing for simplicity.
 
The subtraction term divided by the Born cross section  ${\hat \sigma}_0(Q^2,\epsilon)$\footnote{The Born cross section ${\hat \sigma}_0(Q^2,\epsilon)$ is defined in Ref. \citen{rf:7}.} for the branching process  
\begin{eqnarray}
q(p_q) \rightarrow q(r)+g(p_g),
\end{eqnarray}
 with a subtraction scheme ${\rm F}$  is defined by
 \begin{eqnarray}
 {d{\tilde S}_{\rm NS}^{[\rm F]} \over dz d(-r^2)}={\alpha_s \over 2\pi}{1 \over \Gamma(1-\epsilon)}\left[{-r^2 \over 4\pi\mu^2}\right]^{-\epsilon}{\left({\hat P}_{qq}^{(0)}(z)-\epsilon {\hat Q}_{\rm NS}^{[\rm F]}(z)\right)_+ \over -r^2}
 \end{eqnarray}
 in $4-2\epsilon$ dimensions, where
\begin{eqnarray}
  {\hat P}_{qq}^{(0)}(z)=C_F{1+z^2 \over 1-z}, 
\end{eqnarray}
and ${\hat Q}_{\rm NS}^{[\rm F]}(z)$ depends on the subtraction scheme.\footnote{We define 
$$
 \left({\hat f}(z)\right)_+={\hat f}(z)-\delta(1-z)\int^1_0dy{\hat f}(y) 
$$
for a function ${\hat f}(z)$ unregulated at $z=1$. } 
  Here, $C_F=4/3$ is the color factor.
The strong coupling constant is defined by $\alpha_s\mu^{2\epsilon}$ for the dimensionless coupling $\alpha_s$ and a mass parameter $\mu$.
Here, ${\hat Q}_{\rm NS}^{[\overline{\rm MS}]}(z)=0$ corresponds to the $\overline{\rm MS}$ subtraction scheme. 
The momentum of the quark, $r$, is described in terms of the momentum fraction $z$, the virtuality $r^2$, and the transverse momentum $r_T$ as $ r=p_q-p_g=zp_q+(r^2/{\hat s})p_{\bar q}+r_T$ with $p_q\cdot r_T=p_{\bar q}\cdot r_T=0$. Here, we set $p_q^2=p_{\bar q}^2=p^2_g=0$, because the relations $-p_q^2,-p_{\bar q}^2,p^2_g \ll -r^2$ are expected in parton shower generation.

The regularization of the infrared singularity is defined so as to conserve the particle number of the initial-state quarks for the branching process presented in Eq. (2$\cdot$2).
 
The parton evolution depends on the subtraction scheme for the mass singularity, which appears owing to collinear parton production.  
     Including the NLL order term, the infrared regulated splitting function with the subtraction scheme ${\rm F}$, denoted by $\left(P_{\rm NS}^{[{\rm F}]}(\alpha_s,z)\right)_+$, is defined by \cite{rf:2}
\begin{eqnarray}
  \left({\hat P}_{\rm NS}^{[{\rm F}]}(\alpha_s,z)\right)_+ = {\alpha_s \over 2\pi}\left({\hat P}^{(0)}_{qq}(z)\right)_+ 
    +  \left({\alpha_s \over 2\pi}\right)^2\left({\hat P}_{\rm NS}^{[\overline{\rm MS}](1)}(z)-{\beta_0 \over 2}{\hat Q}_{\rm NS}^{[{\rm F}]}(z)\right)_+   
  \end{eqnarray}
with $\beta_0=11-2/3N_f$ for $N_f$ active flavors.
Here, $\left({\hat P}_{\rm NS}^{[\overline{\rm MS}](1)}(z)\right)_+$ is the NLL order splitting function calculated with the $\overline{\rm MS}$ scheme\cite{rf:13}. 
\footnote{We neglect the process $q \rightarrow {\bar q}+X$, since the contribution is small at the NLL order. } 

The splitting function $\left({\hat P}_{\rm NS}^{[{\rm F}]} (\alpha_s,z)\right)_+$ satisfies 
\begin{eqnarray}
   \int^{1}_{0} dz \left({\hat P}_{\rm NS}^{[{\rm F}]}(\alpha_s,z)\right)_+ = 0
\end{eqnarray}
in each branching process.  Using 
\begin{eqnarray}
\int^1_{1-\delta}dz\left({\hat P}_{\rm NS}^{[{\rm F}]}(\alpha_s,z)\right)_+=-\int^{1-\delta}_0dz{\hat P}_{\rm NS}^{[{\rm F}]}(\alpha_s,z),
\end{eqnarray}
 the nonbranching probablitiy for the nonsinglet quarks is defined by
\begin{eqnarray}
\Pi_{NB}^{[{\rm F}]({\rm NS})} ( K^2_2, K^2_1 ) = {\rm exp} \left[ - \int^{K^2_2 }_{K^2_1} {dK^2 \over K^2 } \int^{1 - \delta}_0 dz{\hat P}_{\rm NS}^{[{\rm F}]} (\alpha_s,z)  \right]. 
\end{eqnarray}
Here, the partons inside an initial-state hadron have spacelike virtualities (i.e., $k_i^2 \equiv -K_i^2 <0$), and $\delta$ denotes a resolution of the momentum fraction of the final-state partons.\footnote{In Monte Carlo calculation, a cutoff parameter for small $z$ integration is chosen, so that a very small ${\hat s}$ for a given hard process is not generated by parton showers.}
    The actual steps in the Monte Carlo method are similar to those  presented in Refs. \citen{rf:3} and \citen{rf:4}.

 Using this algorithm, the particle number distribution of the nonsinglet quarks is reproduced up to their normalization without the introduction of any nontrivial weight factor. 
Therefore, the model guarantees the particle number for the initial-state valence quarks, namely,
\begin{eqnarray}
    \int^1_0dx f_{q_V/p}^{[{\rm F}]}(x,K^2)=N_q 
\end{eqnarray}
is satisfied for any $ K^2 $, where $x$ is the momentum fraction of the valence  quark inside a proton, with $N_u=2$ for the valence $u$ quarks and $N_d=1$ for the valence $d$ quark.  Here, $ f_{q_V/p}^{[{\rm F}]}(x,K^2) $ is the particle number distribution function of the valence  quarks inside a proton calculated with a factorization scheme ${\rm F}$. The above method guarantees universal (process-independent) parton distributions with a factorization scheme $F$, since the same scheme has been implemented in a deep-inelastic process.\cite{rf:14}

\subsection{Kinematical constraint on the hard process}

In Monte Carlo calculation, kinematical constraints must be taken into account,  for the momentum conservation of partons in each generated event.  
Here, the kinematical constraint due to the squared momentum of the photon $Q^2=(r+p_{\bar q})^2$ to $O(\alpha_s)$ accuracy of QCD is imposed in the delta function, in which the quantity ${\hat \tau}=Q^2/{\hat s}$ is given by ${\hat \tau}=z+r^2/{\hat s}$ for on-shell gluon radiation ($p_g^2=0$) (denoted by the $F'$ scheme) as
 \begin{eqnarray}
 {dS_{\rm NS}^{[\rm F']} \over d{\hat \tau}d(-r^2)}=\int^1_0dz{d{\tilde S}_{\rm NS}^{[\rm F]} \over dz d(-r^2)}\delta(z-{\hat \tau}-(-r^2)/{\hat s}).
 \end{eqnarray}
We define the integrated contribution as 
\begin{eqnarray}
 \int_0^{{\hat s}w^{[\rm F'](I)}} d(-r^2){dS_{\rm NS}^{[\rm F']} \over d{\hat \tau}d(-r^2)}= {\alpha_s \over 2\pi}{1 \over \Gamma(1-\epsilon)}\left[{{\hat s} \over 4\pi\mu^2}\right]^{-\epsilon}{\tilde F}_{\rm NS}^{[F'](I)}(\epsilon,{\hat \tau}),
 \end{eqnarray}
 where $I$ denotes the region of phase space for the hard scattering process being considered. Here, ${\hat s}w^{[F'](I)}$ is a limit of the $-r^2$ integration.

 According to Eq. (2$\cdot$10), the phase space $0 \leq -r^2 \leq M^2$ with $z \leq 1$ for the initial-state parton radiation  covers the region  $0 \leq -{\hat t} \leq (1-{\hat \tau}){\hat s}$ for $1-{\hat \tau}_M \leq {\hat \tau} \leq 1(I=S)$ and $0 \leq -{\hat t} \leq M^2$ for ${\hat \tau} \leq 1-{\hat \tau}_M(I=C)$ in the hard scattering cross section, with ${\hat \tau}_M \equiv M^2/{\hat s}$.  The collinear contribution is also subtracted from the antiquark leg.
  Here, we consider the two regions $1-{\hat \tau}_M \leq {\hat \tau} \leq 1$ and ${\hat \tau} \leq 1-{\hat \tau}_M$ separately for the subtraction terms. 

For $1-{\hat \tau}_M \leq {\hat \tau} \leq 1$, we subtract the collinear contribution from the hard scattering cross section in the range $0 \leq -{\hat t} \leq {\hat s}(1-{\hat \tau}) $.   
 If we ignore the term $r^2/{\hat s}$ in the delta function in Eq. (2$\cdot$10) and subtract the collinear contribution in the range $ 0 \leq -r^2 \leq M^2$, with ${\hat \tau}=z$, the matching between the hard scattering cross section and the initial-state radiation is broken.  In this case, $M^2$ is larger than the kinematical boundary for $-{\hat t}(=-r^2)$, given by ${\hat s}(1-{\hat \tau})$ in the region $1-{\hat \tau}_M \leq {\hat \tau} \leq 1$.

\subsection{Kinematical constraint on the branching processes}

  The algorithm for the generation of transverse momenta in the initial-state radiation is similar to that for the Monte Carlo model presented in  the LL order of QCD,  except that the effects due to the three-body decay functions, which are the $O(\alpha_s^2)$ contributions,  are taken into account for  each branching step.

In the flavor nonsinglet sector, the contribution of the two-gluon radiation, such as
\begin{eqnarray}
 q(k_0) \rightarrow g(l_1)+g(l_2)+q(k) ,
\end{eqnarray}
becomes large in the soft gluon region.
Here, the momenta of these partons are  denoted by $k_0$, $l_1,l_2$ and $k$, with $l_1^2=l_2^2=0$ and $k_0^2,k^2 <0$.

 In our algorithm, the transverse momenta of partons were generated according to  the branching probability restricted by the boundary condition determined by the NLL order terms.  In the case of Eq. (2$\cdot$12), the transverse momentum of the quark with momentum $k$ is given by
 \begin{eqnarray}
  {\vec k_T}^2=z(1-z)\left[ k_0^2 + {-k^2 \over z} - {l^2 \over 1-z}\right]. 
   \end{eqnarray} 
Here, we define $l^2=(l_1+l_2)^2$ and $z$ denotes the longitudinal momentum fraction of the parton momentum $k$ for the momentum $k_0$.

The upper limit of the virtuality $l^2$ for the radiated partons is not properly determined at the LL order, since the kinematical constraint of the branching process is a part of the NLL order contribution.  As shown in Ref.\citen{rf:4}, the boundary condition is determined by the parton shower algorithm as well as the NLL order terms.

  For strong ordered virtuality  of the initial-state partons, namely, $-(k_0-l_1)^2,-(k_0-l_2)^2 < -k^2$, the virtuality $l^2$ is restricted by the condition $l^2 \leq f_{\rm NS}^{[\rm F]}(-k^2)$ at the NLL order, instead of $l^2 \leq (1-z)/z(-k^2)$, by the kinematic boundary of the two-body branching, due to the fact that ${\vec k_T}^2 \geq 0$ for $-k^2_0 \ll -k^2 , l^2$ at the LL order approximation.  
Here, the factor of $f_{\rm NS}^{[\rm F]}$ is given by the NLL terms calculated with the subtraction scheme $F$.

 In the case considered above, the two-body branching process is allowed in the branching steps of the initial-state parton radiations, as is usually the case in LL order parton shower models, except that the contributions due to the NLL order terms for the initial-state radiations are included in the kinematical boundary for the radiated partons.   

  \subsection{Factorized cross sections}

  The hard scattering cross section with the factorization scheme ${\rm F}'$ is defined by
\begin{eqnarray}
{d{\hat \sigma}_{q{\bar q}(NS)}^{[{\rm F}'](I)} \over d{\hat \tau}}=
{\alpha_s \over 2\pi}{\hat \sigma}_0(Q^2,0)K_{q{\bar q}(NS)}^{[{\rm F}'](I)}({\hat \tau})
\end{eqnarray}
with 
\begin{eqnarray}
{\hat \sigma}_0(Q^2,0)={4\pi\alpha^2 \over 3N_CQ^2}{\hat e}_q^2.
\end{eqnarray}
 Here, $N_C=3$ and the electric coupling constant of the quark is defined by ${\hat e}_q^2\alpha$.

\subsubsection{Soft gluon region}

For the soft gluon region $1-2{\hat \tau}_M \leq {\hat \tau} \leq 1$, we consider two separate regions, one for $1-{\hat \tau}_M \leq {\hat \tau} \leq 1~ (I=S1)$ and the other for $1-2{\hat \tau}_M \leq {\hat \tau} < 1-{\hat \tau}_M ~ (I=S2)$. 

For $I=S1$, we have 
\begin{eqnarray}
K_{q{\bar q}(NS)}^{[{\rm F}'](S1)}({\hat \tau})= K_{q{\bar q}}({\hat \tau},\epsilon)-2{\tilde F}^{[{\rm F}'](S)}_{NS}(\epsilon,{\hat \tau}),
\end{eqnarray}
where $K_{q{\bar q}}({\hat \tau},\epsilon)$ denotes the DY cross section integrated over $0 \leq -{\hat t} \leq (1-{\hat \tau}){\hat s}$.\cite{rf:7,rf:15}

For $I=S2$, we have
\begin{eqnarray}
K_{q{\bar q}(NS)}^{[{\rm F}'](S2)}({\hat \tau})= K_{q{\bar q}}({\hat \tau},\epsilon)-2{\tilde F}^{[{\rm F'}](C)}_{NS}(\epsilon,{\hat \tau}),
\end{eqnarray}
since $M^2 \leq {\hat s}(1-{\hat \tau})$ in this region.

\subsubsection{Hard collinear region}

The remnant of the collinear gluon contribution of the hard scattering cross section integrated over the range $0 \leq -{\hat t} \leq M^2$ ( or  $0 \leq -{\hat u} \leq M^2$)  for $ {\hat \tau} < 1-2{\hat \tau}_M~(I=C)$ is given by
\begin{eqnarray}
K_{q{\bar q}(NS)}^{[{\rm F'}](C)}({\hat \tau})= K_{q{\bar q}}^{(C)}({\hat \tau},\epsilon)-{\tilde F}^{[{\rm F}'](C)}_{\rm NS}(\epsilon,{\hat \tau}),
\end{eqnarray}
where $K_{q{\bar q}}^{(C)}({\hat \tau},\epsilon)$ denotes the DY cross section integrated over $0 \leq -{\hat t} \leq M^2$ ( or  $0 \leq -{\hat u} \leq M^2$).\cite{rf:7}

 \section{ Numerical results in the ${\rm  JC}$ scheme}
 
 In this section, some numerical results obtained using the algorithm explained in the previous section are presented.  
In the parton shower model explained in the previous section, the virtuality of the initial-state partons is generated with strongly ordered configuration, namely, $M_0^2 \leq K^2_1 \leq K^2_2 \leq \cdots, \leq M^2$. It has been discussed that the strongly ordered configuration of the parton virtuality specifies a factorization scheme,\cite{rf:12} which we refer to as the ${\rm JC}$ scheme.  Therefore, according to the Monte Carlo algorithm for the initial-state radiations, we implement the ${\rm JC}$ scheme explained in Refs. \citen{rf:8} and \citen{rf:10}, where we define 
\begin{eqnarray}
{\hat Q}_{\rm NS}^{[\rm JC]}(z)={\hat P}^{(0)}_{qq}(z)\log(1-z)-{\hat P}'_{qq}(z)
\end{eqnarray}
in Eq. (2$\cdot$3).  Here, the function ${\hat P}'_{qq}(z)=-C_F(1-z)$ is an $O(\epsilon)$ term of the splitting function in $4-2\epsilon$ dimensions.\footnote{The splitting function in $4-2\epsilon$ dimensions is defined by \cite{rf:16}
\begin{eqnarray*}
{\hat P}_{qq}(z,\epsilon)=  {\hat P}_{qq}^{(0)}(z)+\epsilon {\hat P}'_{qq}(z).
\end{eqnarray*}
}

Since  $K^{(H)}_{q{\bar q}}({\hat \tau}) \gg K_{q{\bar q}(NS)}^{[{\rm JC}'](C)}({\hat \tau}) $ for ${\hat \tau} < 1-2{\hat \tau}_M$, as presented in Ref. \citen{rf:8}, the remnant of the hard collinear region after subtracting the collinear contribution with the ${\rm JC}'$ scheme can be safely neglected even at the NLL order accuracy.
Here, $K^{(H)}_{q{\bar q}}({\hat \tau})$ is the DY cross section for hard gluon radiation presented in Eq. (2$\cdot$1) integrated over $M^2 \leq -{\hat t} \leq (1-{\hat \tau}){\hat s}-M^2$, and $K_{q{\bar q}(NS)}^{[{\rm JC}'](C)}({\hat \tau})$ is the same as  that of the singlet sector given in Ref. \citen{rf:8}.

The Monte Carlo method for the Drell-Yan process that we used in this paper is presented in Appendix A. The explicit expressions of the factorized cross sections in the ${\rm JC}'$ scheme implemented in  numerical calculations are shown in Appendix B.

 Here, we present the transverse momentum distributions of the virtual photon $\gamma^*$ produced by a $u$-quark and an anti-$u$-quark annihilation in $p{\bar p}$ collisions with the center of mass energy at $\sqrt{s}=1800~{\rm GeV}$.
  
\subsection{Ambiguities at the LL order approximation}

First, we investigate the ambiguities at the LL order approximation of QCD.

In Monte Carlo models based on the LL order of QCD, the initial-state parton showers are generated using the LL order splitting function, which has no compensation term for a subtraction method in the hard scattering cross section as well as the kinematical boundary for the hard scattering process.

In the Monte Carlo calculation, the parton showers  start from $M_0^2=5~{\rm GeV}^2$ in both sides of protons. After the evolutions are terminated, the momenta of the generated spacelike partons $p_q$ and $p_{\bar q}$ are reconstructed. Then, the $s$-channel momentum is calculated as ${\hat s}=(p_q+p_{\bar q})^2=Q^2$. 

At the LL order of QCD, the upper limit of the virtuality for the generated parton is restricted by a large energy scale $M$, which can be chosen arbitrarily. 
 In Fig. 1, the transverse momentum distributions for the virtual photon integrated over the range of the photon virtuality  $40~{\rm GeV} \leq Q \leq 60~{\rm GeV}$ are represented  by the solid histogram for $M=40~{\rm GeV}$ and by the dashed histogram for $M=60~{\rm GeV}$ with $l^2=l_0^2=0.5~{\rm GeV}^2$, where the two-body decay kinematics shown in Eq. (2$\cdot$13) for initial-state parton radiations are imposed.  
 The transverse momentum distribution clearly depends on the choice of $M$. Furthermore, the contributions only from  the initial-state radiations are insufficient to reproduce the distribution in the hard $q_T$ region. Therefore, the NLO contribution should be added.

Another ambiguity comes from the kinematical boundary in the initial-state parton branching vertices explained in 2.3.
  In Fig. 1, the dashed histogram shows a case with $\l_0^2 \leq l^2 \leq (1-z)/z(-k^2)$ due to $k_T^2 \geq 0$ in Eq. (2$\cdot$13) at the LL order approximation for $M=40~{\rm GeV}$.\footnote{The virtuality $l^2$ for outgoing partons is generated by using the nonbranching probability for timelike parton evolutions. } 

In Fig. 1, the transverse momenta according to the cross section at the NLO contributions are also presented for $M=40~{\rm GeV}$ by the plus symbols and for $M=60~{\rm GeV}$ by the crossed symbols, which also depend on the factorization scale $M$.

\begin{figure}
\centerline{\includegraphics[width=10cm]{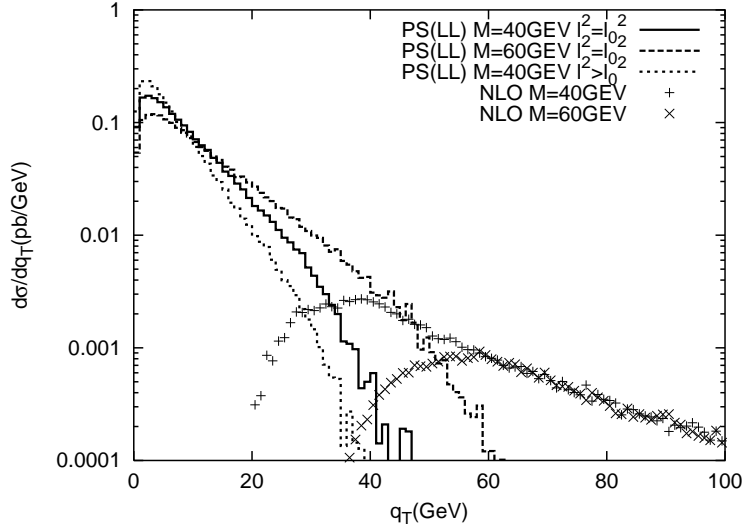}}

\caption{Transverse momentum distributions for the virtual photon integrated over $40~{\rm GeV} \leq Q \leq 60~{\rm GeV}$ with $\sqrt{s}=1800~{\rm GeV}$.  The solid and dashed histograms represent the results for $M=40~{\rm GeV}$ and $M=60~{\rm GeV}$ with $l^2=l_0^2=0.5~{\rm GeV}^2$, respectively.  The dashed histogram shows a case with $\l_0^2 \leq l^2 \leq (1-z)/z(-k^2)$ for $M=40~{\rm GeV}$. The NLO contributions are represented for $M=40~{\rm GeV}$ by the plus symbols and for $M=60~{\rm GeV}$ by the crossed symbols.
}
\end{figure}
\begin{figure}
\centerline{\includegraphics[width=10cm]{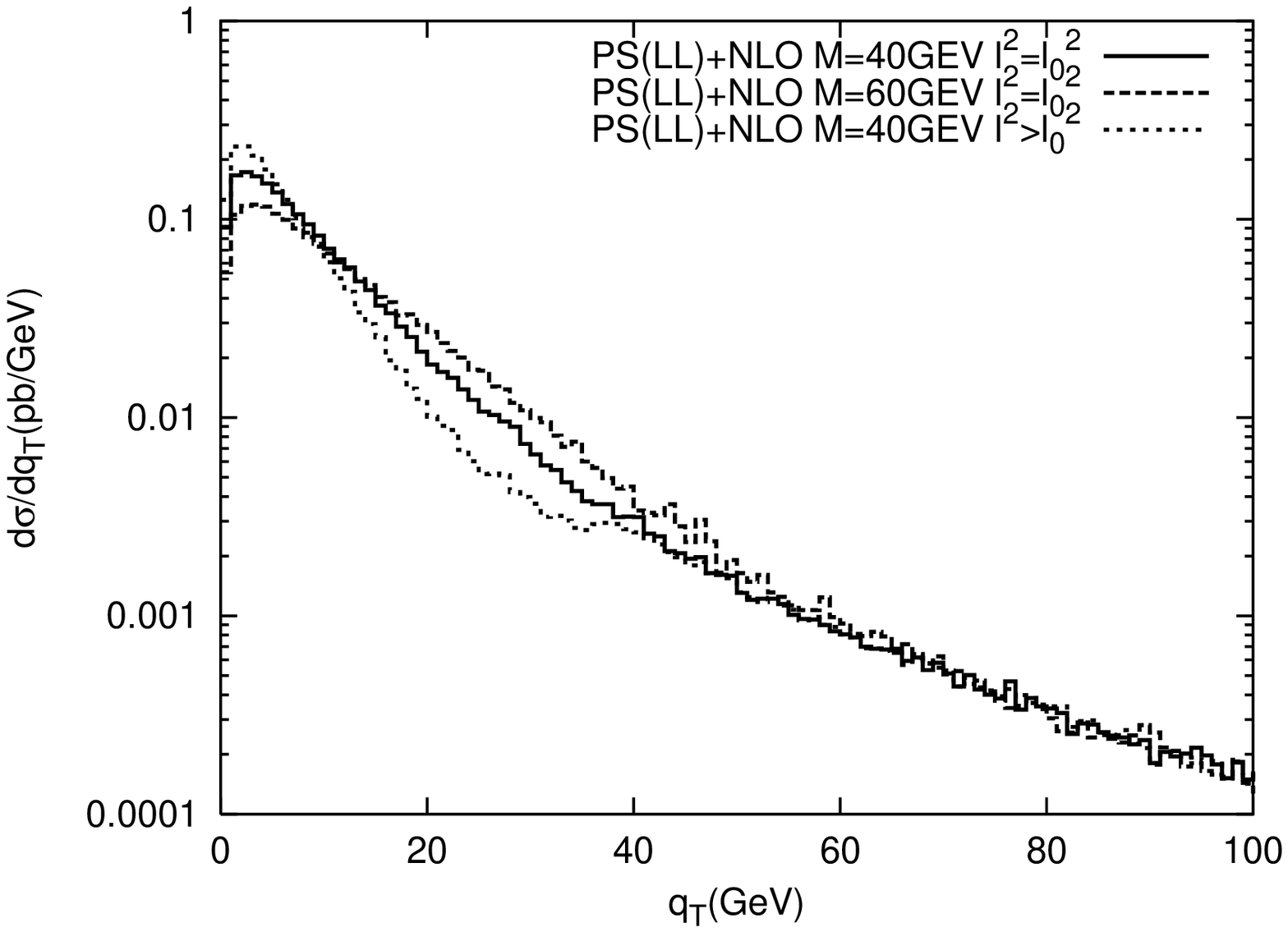}}
\caption{Transverse momentum distributions for the virtual photon integrated over $40~{\rm GeV} \leq Q \leq 60~{\rm GeV}$ with $\sqrt{s}=1800~{\rm GeV}$.  The solid and dashed histograms represent the results for $M=40~{\rm GeV}$ and  $M=60~{\rm GeV}$ with $l^2=l^2_0=0.5~{\rm GeV}^2$, respectively. The dotted histogram represents the result for $\l_0^2 \leq l^2 \leq (1-z)/z(-k^2)$ with $M=40~{\rm GeV}$. 
}
\end{figure}

 The transverse momentum distributions combined with the initial-state radiations and the NLO contributions are presented in Fig. 2.  Although the $M$ dependence of the transverse momentum distributions due to the initial-state radiations is partially compensated by that from the NLO contribution,  the calculated results at the LL order approximation still depend on the factorization scale $M$ as well as the phase space boundary for the branching vertices.

 \subsection{The NLL order contributions}

The transverse momentum distributions including the NLL order contributions, explained in $\S$2, are presented in Fig. 3. Here, the NLO contribution with the ${\rm JC}'$ scheme is implemented, where the phase space boundary of the hard scattering cross sections is imposed in subtraction terms as well as the generation of the initial-state parton radiations. 
 
Furthermore, the initial-state radiations are generated within the phase space, which satisfies ${\hat s}_{m,n}-{\hat s}_{m+1,n} >-p_{q(m+1)}^2~({\hat s}_{m,n}-{\hat s}_{m,n+1} >-p_{{\bar q}(n+1)}^2)$ with ${\hat s}_{m,n}=(p_{q(m)}+p_{{\bar q}(n)})^2$ for each branching step. Here, $p_{q(m)}(p_{{\bar q}(m)})~(m=0,1,\cdots)$ denotes the momentum of the initial-state quark (antiquark) generated in the $m$-th branching step in the forward evolution algorithm.\footnote{The parton evolutions start from the distribution function at $M_0$, which is approximately given by\cite{rf:10}
\begin{eqnarray*}
f^{[\rm JC]}_{u_V/p}(x,M_0^2) \simeq f^{[\overline{\rm MS}]}_{u_V/p}(x,M_0^2)+{\alpha_s(M_0^2) \over 2\pi}\int^1_x{dz \over z}\left({\hat Q}_{\rm NS}^{[\rm JC]}(z)\right)_+ f^{[\overline{\rm MS}]}_{u_V/p}\left({x \over z},M_0^2\right)
\end{eqnarray*}
with the GRV(98) distribution function for $ f^{[\overline{\rm MS}]}_{u_V/p}(x,M_0^2)$.\cite{rf:17}
}

In Fig. 3, the transverse momentum distributions for the virtual photon integrated over the range of the photon virtuality  $40~{\rm GeV} \leq Q \leq 60~{\rm GeV}$ are presented by the solid histogram for $M=40~{\rm GeV}$ and by the dashed histogram for $M=60~{\rm GeV}$.  By including the NLL order contributions, ambiguities due to the choice of the factorization scale become sufficiently small compared with the case at the LL order of QCD.  An analytic calculation of the transverse momentum distribution for one-gluon radiation cross section multiplied by $K=\exp(\alpha_s(Q^2)/2\pi C_F\pi^2)$ integrated over the range $40~{\rm GeV} \leq Q \leq 60~{\rm GeV}$ is also shown by the dotted curve. The generated cross section approximately reproduces the enhancement due to the next-to-next-to-leading order (NNLO) corrections ($O(\alpha_s^2$) terms) for the Drell-Yan cross section in the large $q_T$ region. \cite{rf:18}

Here, the transverse momenta of the initial-state partons with spacelike virtuality are generated using the effective two-body vertices, where the three-body decay functions are included as the boundary conditions, namely, $l_0^2 \leq l^2 \leq f_{\rm NS}^{[\rm JC]}(-k^2)$, for the virtuality $l^2$ of the outgoing partons in Eq. (2$\cdot$13), with  
\begin{eqnarray}
f_{\rm NS}^{[\rm JC]}(z_1,z_2,z)=\exp\left[{V_{qgg}(z_1,z_2,z) \over L^{[A]}(z_1,z_2,z)}\right].
\end{eqnarray}
Here, the function $L^{[A]}$ is the coefficient of the mass singular term given by 
\begin{eqnarray}
L^{[A]}(z_1,z_2,z)={{\hat P}_{qq}^{(0)}(z) \over z_1+z_2}{\hat P}_{gg}^{(0)}\left({z_1 \over z_1+z_2}\right)
\end{eqnarray}
with $z_1+z_2+z=1$, for the process  $q(k_0) \rightarrow g(l_1+l_2)+q(k) \rightarrow g(l_1)+g(l_2)+q(k)$, where  $P_{gg}^{(0)}(z)$ is the LL order splitting function for the $g \rightarrow gg$ process, and the function $V_{qgg}(z_1,z_2,z)$ is given in Ref. \citen{rf:14}.  Here, $z_i~(i=1,2)$ denotes the momentum fraction of parton momentum $l_i$ for the momentum $k_0$.

\begin{figure}
\centerline{\includegraphics[width=10cm]{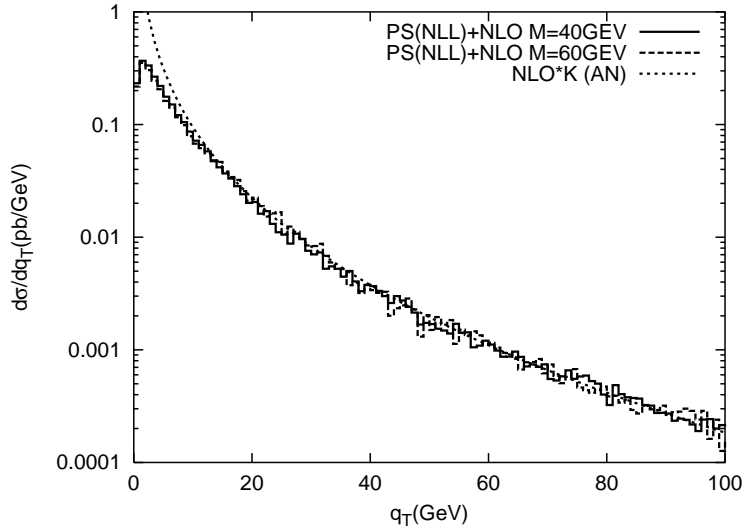}}
\caption{Transverse momentum distributions for the virtual photon integrated over $40~{\rm GeV} \leq Q \leq 60~{\rm GeV}$ with $\sqrt{s}=1800~{\rm GeV}$, where the NLL order contributions are taken into account.  The solid and dashed histograms represent the results for $M=40~{\rm GeV}$ and $M=60~{\rm GeV}$ obtained with the ${\rm JC}'$ scheme, respectively. The dotted curve represents the one-gluon radiation cross section multiplied by $K=\exp(\alpha_s(Q^2)/2\pi C_F\pi^2)$ integrated over $40~{\rm GeV} \leq Q \leq 60~{\rm GeV}$.   
}
\end{figure}

 The asymptotic form for the soft gluon radiation with the above choice gives $f_{\rm NS}^{\rm [JC]}\sim z_1$ for $z_1 \ll z_2,z$,\cite{rf:4} which corresponds to the angular ordering condition, such as  $\theta_{l_1l_2}<\theta_{k_0l_2}$\cite{rf:14,rf:19,rf:20} for the  small angle approximation, which is the case for $\theta_{l_1l_2},\theta_{k_0l_2} \ll 1$.\footnote{Apart from the accuracy of  the scaling  violation generated by parton showers, the choice of the upper limit as $l^2 \le (-k^2) z_1$ for $z_1 \ll z_2,z$, which is a frame-independent form compared with the angular ordering condition,  may be useful even at the LL order approximation in parton evolutions.  In this choice,  the ambiguity due to the kinematical boundary for the branching vertices is approximately eliminated. }

\section{Summary and comments}

In this work, we have studied the Monte Carlo algorithm in hadron-hadron collisions, in which the initial-state radiation is generated using a parton shower model including the next-to-leading logarithmic (NLL) order contributions. 

We calculated hard scattering cross sections for the process $q{\bar q} \rightarrow \gamma^*g$ with the jet calculus scheme, in which a kinematical constraint for the gluon radiation is taken into account. This method is called the ${\rm JC}'$ scheme.  In this method, this kinematical constraint guarantees a proper phase space boundary for the subtraction terms.

At the leading-logarithmic (LL) order approximation, we have shown that the transverse momentum distributions of the virtual photon depend on the factorization scale even the next-to-leading order (NLO) contribution in the hard scattering cross section is added in the large transverse region.   We also pointed out that the transverse momentum distributions depend on the kinematical boundary of the branching processes in the initial-state parton radiations, which cannot be determined at the LL order approximation.

The kinematical conditions of the parton radiations and the remnant of the collinear region, in which the mass singularity is subtracted, are parts of the NLL order contributions. Therefore, these terms cannot be neglected at the NLL order accuracy.  In the algorithm presented in this paper, these contributions are included in Monte Carlo calculations.

Furthermore, the jet calculus  scheme provides a matching between the initial-state radiations and the hard scattering cross sections.  
By including the NLL order contributions, we found that ambiguities due to the choice of the factorization scale become small compared with the case at the LL order of QCD.  Furthermore, the kinematical boundary for the branching vertices, which is not determined at the LL order of QCD, is given by the three-body decay functions ($O(\alpha_s^2)$ contributions for the branching vertex).
 
For the large $q_T$ region, the cross section is enhanced, which is consistent with the K-factor $\simeq \exp(\alpha_s/2\pi C_F\pi^2)$ for the Drell-Yan processes.  The generated cross section approximately reproduces the results including the  $O(\alpha_s^2)$ contributions for the hard scattering cross sections.\cite{rf:18}

The method presented in this paper  may be useful for the construction of more accurate Monte Carlo algorithms, in which the parton radiation at the NLL order as well as the hard scattering cross sections at the NLO of QCD are taken into account. 

 This method can be extended to the flavor singlet sector implementing the parton shower algorithm based on momentum distributions studied in Ref. \citen{rf:10}.

It may be important to evaluate various processes by using Monte Carlo methods including the NLL order contributions with the NLO terms, such as $W$+jet and $t{\bar t}$+jet processes, which are backgrounds for new physics research as well as Higgs boson productions in hadron-hadron collisions.

\section*{Acknowledgements}

This work was partially supported by the Research Center for Measurement in Advanced Science of Rikkyo University, and in part by the Rikkyo University Special Fund for Research.


\vspace{5mm}

\begin{center}
{\Large Appendix A}
\end{center}
\vspace{5mm}


In Monte Carlo calculation of hard scattering contributions, we generate the photon virtuality $Q^2$ and the transverse momentum ${\vec q}_T$ with appropriate weight according to the hard scattering cross section.
  Although the hard scattering cross sections may be evaluated by using more accurate numerical methods,  we implement a rather simple method for testing the algorithm.

In order to evaluate the hard scattering cross section, we integrate Eq. (2$\cdot$14) over the range $b{\hat s} \leq Q^2 \leq a{\hat s}$, with fixed ${\hat s}$.  Here, ${\hat s}$ is generated by the parton showers.  The energy scale of the running coupling constant $\alpha_s$ in this region can be chosen as $Q^2={\hat \tau}{\hat s}$. 
Therefore, the integrated cross section is given by
\begin{eqnarray*}
{\hat \sigma}_{q{\bar q}(NS)}^{[{\rm JC}'](I)} (a,b)=\int^a_b d{\hat \tau}{d{\hat \sigma}_{q{\bar q}(NS)}^{[{\rm JC}'](I)} \over d{\hat \tau}}=\int^a_b d{\hat \tau}{\alpha_s({\hat \tau}{\hat s}) \over 2\pi}{\hat \sigma}_0({\hat \tau}{\hat s},0)K_{q{\bar q}(NS)}^{[{\rm JC}'](I)}({\hat \tau}).
\end{eqnarray*}

For the soft gluon region $1-2{\hat \tau}_M \leq {\hat \tau} \leq 1$, we consider two separate regions, one for $1-{\hat \tau}_M \leq {\hat \tau} \leq 1~ (I=S1)$ and the other for $1-2{\hat \tau}_M \leq {\hat \tau} < 1-{\hat \tau}_M ~ (I=S2)$. 

For $I=S1$, we have 
\begin{eqnarray*}
K_{q{\bar q}(NS)}^{[{\rm JC}'](S1)}&&({\hat \tau})
=C_F\Big[-4\left({\log(1-{\hat \tau}) \over 1-{\hat \tau}}\right)_+-{3 \over (1-{\hat \tau})_+} \nonumber  \\
& & + \left(-{7 \over 2}+{4 \over 3}\pi^2\right)\delta(1-{\hat \tau})\Big].
\end{eqnarray*}
The factorized cross section has no $M$ dependence, because the subtraction term is integrated over the range $0 \leq -r^2 \leq {\hat s}(1-{\hat \tau})$ due to the delta function in Eq. (2$\cdot$10). 

Here,
\begin{eqnarray*}
K_{q{\bar q}(NS)}^{[{\rm JC}'](S1)}({\hat \tau})=K_{q{\bar q}}^{[{\rm JC}'](S)}({\hat \tau})-C_F\delta(1-{\hat \tau}),
\end{eqnarray*}
because the subtraction term is given by  
\begin{eqnarray*}
{\tilde F}_{NS}^{[{\rm JC}'](S)}(\epsilon,{\hat \tau})={\tilde F}_{qq}^{[{\rm JC}'](S)}(\epsilon,{\hat \tau})+{C_F \over 2}\delta(1-{\hat \tau}),
\end{eqnarray*}
where $K_{q{\bar q}}^{[{\rm JC}'](S)}({\hat \tau})$ and ${\tilde F}_{qq}^{[{\rm JC}'](S)}(\epsilon,{\hat \tau})$ are given in Ref. \citen{rf:8} for the singlet sector with infrared regularization so that the total momentum of partons in the initial-state is conserved.

For $I=S2$, we have
\begin{eqnarray*}
K_{q{\bar q}(NS)}^{[{\rm JC}'](S2)}&&({\hat \tau})
=2C_F\Big[{2 \over 1-{\hat \tau}}\log{1-{\hat \tau}-{\hat \tau}_M \over {\hat \tau}_M}-(1+{\hat \tau})\log{1-{\hat \tau} \over {\hat \tau}_M} \nonumber \\
& & -(1-{\hat \tau}-{\hat \tau}_M)\Big],
\end{eqnarray*}
since $M^2 \leq {\hat s}(1-{\hat \tau})$ in this region.
Here, the subtraction term in Eq. (2$\cdot$17) is 
\begin{eqnarray*}
{\tilde F}_{NS}^{[{\rm JC}'](C)}(\epsilon,{\hat \tau})={\tilde F}_{qq}^{[{\rm JC}'](C)}(\epsilon,{\hat \tau}),
\end{eqnarray*}
where ${\tilde F}_{qq}^{[{\rm JC}'](C)}(\epsilon,{\hat \tau})$ is given in Ref. \citen{rf:8}. 

In order to evaluate the soft gluon contribution, we integrate over the range $(1-\eta_s) \leq {\hat \tau} \leq 1$, with fixed ${\hat s}$. Here, $\eta_s$ is a cutoff parameter satisfying $\eta_s \leq 2{\hat \tau}_M$ with ${\hat \tau}_M=M^2/{\hat s}$.

In this model, the soft gluon contributions, $(I=S)$, are resummed as 
\begin{eqnarray*}
{\hat \sigma}_0({\hat s},0)+{\hat \sigma}_{q{\bar q}(NS)}^{[{\rm JC}'](S)} (1,1-\eta_s)\simeq{\hat \sigma}_0({\hat s},0)R_{q{\bar q}(NS)}^{[{\rm JC}'](S)} ({\hat s},\eta_s) 
\end{eqnarray*} 
with
\begin{eqnarray*}
R_{q{\bar q}(NS)}^{[{\rm JC}'](S)}({\hat s},\eta_s) =\exp\left[{\hat \sigma}_{q{\bar q}(NS)}^{[{\rm JC}'](S)} (1,1-\eta_s)/{\hat \sigma}_0({\hat s},0)\right].
\end{eqnarray*}

The calculated results for the range $1-\eta_s \leq {\hat \tau} \leq 1$ are given by 
\begin{eqnarray*}
{\hat \sigma}_{q{\bar q}(NS)}^{[{\rm JC}'](S)} (1,1-\eta_s)={\hat \sigma}_{q{\bar q}(NS)}^{[{\rm JC}'](S1)} (1,1-\eta_s)
\end{eqnarray*}
for $0 < \eta_s \leq {\hat \tau}_M$ and  
\begin{eqnarray*}
{\hat \sigma}_{q{\bar q}(NS)}^{[{\rm JC}'](S)} (1,1-\eta_s)={\hat \sigma}_{q{\bar q}(NS)}^{[{\rm JC}'](S1)} (1,1-{\hat \tau}_M)+{\hat \sigma}_{q{\bar q}(NS)}^{[{\rm JC}'](S2)} (1-{\hat \tau}_M,1-\eta_s)
\end{eqnarray*}
for ${\hat \tau}_M < \eta_s \leq 2{\hat \tau}_M$.

For ${\hat \tau}_0 \geq1-2{\hat \tau}_M ({\hat \tau}_0 =Q_0^2/{\hat s}$), we set $Q^2={\hat s}$ and ${\vec q}_T=({\vec p}_q+{\vec p}_{\bar q})_T$ with the weight given by ${\hat \sigma}_0({\hat s},0)R_{q{\bar q}(NS)}^{[{\rm JC}'](S)}({\hat s},1-{\hat \tau}_0)$. Here, $Q_0^2$ is the minimum virtuality of photons to be generated.

For ${\hat \tau}_0 < 1-2{\hat \tau}_M$, we compare a uniformly generated random number $R (0 \leq R \leq 1)$ with the ratio $R_{\sigma}$ defined by
\begin{eqnarray*}
R_{\sigma}={{\hat \sigma}_0({\hat s},0)R_{q{\bar q}(NS)}^{[{\rm JC}'](S)} ({\hat s},2{\hat \tau}_M) \over \sigma_{q{\bar q}(NS)}^{[{\rm JC}']}(1,{\hat \tau}_0) }
\end{eqnarray*}
with 
\begin{eqnarray*}
{\hat \sigma}_{q{\bar q}(NS)}^{[{\rm JC}']}&&(1,{\hat \tau}_0)\simeq {\hat \sigma}_0({\hat s},0)R_{q{\bar q}(NS)}^{[{\rm JC}'](S)} ({\hat s},2{\hat \tau}_M) 
 \Big[1+{\hat \sigma}_{q{\bar q}}^{(H)}(1-2{\hat \tau}_M,{\hat \tau}_0) \\
&& +2{\hat \sigma}_{q{\bar q}}^{[{\rm JC}'](C)}(1-2{\hat \tau}_M,{\hat \tau}_0)\Big].
\end{eqnarray*}

Since ${\hat \sigma}_{q{\bar q}}^{(H)}(1-2{\hat \tau}_M,{\hat \tau}_0) \gg {\hat \sigma}_{q{\bar q}(NS)}^{[{\rm JC}'](C)}(1-2{\hat \tau}_M,{\hat \tau}_0)$, we can safely neglect the remnant of the hard collinear contribution ${\hat \sigma}_{q{\bar q}(NS)}^{[{\rm JC}'](C)}(1-2{\hat \tau}_M,{\hat \tau}_0)$ in the factorized cross section, even at the NLL order accuracy. \cite{rf:8}  On the other hand, in both the $\overline{\rm MS}$ and $\overline{\rm MS}'$ schemes, the remnant of the collinear region becomes negative. Therefore, strong cancellation may occur between the hard scattering contribution and the initial-state radiations.\cite{rf:8}

If $R_{\sigma} > R$, we set $Q^2={\hat s}$ and ${\vec q}_T=({\vec p}_q+{\vec p}_{\bar q})_T$ with the weight given by $\sigma_{q{\bar q}(NS)}^{[{\rm JC}']}(1,{\hat \tau}_0)$.

If $R_{\sigma} < R$, ${\hat \tau}$ is generated  by solving the equation $R={\hat \sigma}_{q{\bar q}(NS)}^{[{\rm JC}']}(1,{\hat \tau})/{\hat \sigma}_{q{\bar q}(NS)}^{[{\rm JC}']}(1,{\hat \tau}_0)$
 for given $R$ with the weight $\sigma_{q{\bar q}(NS)}^{[{\rm JC}']}(1,{\hat \tau}_0)$.
Here, the contribution of the hard gluon radiation ($I=H$) integrated over the range $M^2 \leq -{\hat t} \leq {\hat s}(1-{\hat \tau})-M^2$ is given by 
\begin{eqnarray*}
 K^{(H)}_{q{\bar q}}({\hat \tau})&=&{1 \over {\hat s}}\int^{{\hat s}(1-{\hat \tau})-M^2}_{M^2}d(-{\hat t}){\tilde K}^{(r)}_{q{\bar q}}({\hat \tau},-{\hat t}) \nonumber \\
&=& 2C_F\left[{1 \over C_F} {\hat P}_{qq}^{(0)}({\hat \tau})\log{1-{\hat \tau}-{\hat \tau}_M \over {\hat \tau}_M}-(1-{\hat \tau}-2{\hat \tau}_M)\right] 
\end{eqnarray*}
with
 \begin{eqnarray*}
{\tilde K}^{(r)}_{q{\bar q}}({\hat \tau},-{\hat t})=C_F\left[{\hat s}\left({1 \over -{\hat t}}+{1 \over {\hat s}(1-{\hat \tau})+{\hat t}}\right){1 \over C_F} {\hat P}_{qq}^{(0)}({\hat \tau})-2\right].
\end{eqnarray*}

 The transverse momentum of the photon is constructed as ${\vec q}^2_T=(-{\hat t})({\hat s}+{\hat t}-Q^2)/{\hat s}$, where $-{\hat t}$ is generated according to the distribution of ${\tilde K}^{(r)}_{q{\bar q}}({\hat \tau},-{\hat t})$  
within $M^2 \leq -{\hat t} \leq (1-{\hat \tau}){\hat s}-M^2$.

\vspace{5mm}


\begin{center}
{\Large Appendix B}
\end{center}
\vspace{5mm}

In this appendix, the explicit expressions, which are implemented in the numerical calculations in the text with the ${\rm JC}'$ scheme, are presented. 

In order to simplify our analysis, we evaluate the cross section for the soft gluon radiation with the coupling constant $\alpha_s(Q^2)\simeq\alpha_s({\hat s})$. The running coupling constant for the accuracy of the NLL order is normalized as $\alpha_s(M_Z^2)=0.114$ at the $Z^0$ boson mass.\cite{rf:17}

The integrated cross section over the range $(1-\eta_s){\hat s} \leq Q^2 \leq {\hat s}$ is given by 
\begin{eqnarray*}
\sigma_{q{\bar q}(NS)}^{[{\rm JC}'](S)} (1,1-\eta_s)\simeq{\alpha_s({\hat s}) \over 2\pi}{\hat \sigma}_0({\hat s},0)I_{q{\bar q}(NS)}^{[{\rm JC}'](S)} (1,1-\eta_s)
\end{eqnarray*}
with
\begin{eqnarray*}
I_{q{\bar q}(NS)}^{[{\rm JC}'](S)} (1,1-\eta_s)=\int^1_{1-\eta_s} {d{\hat \tau} \over {\hat \tau}}K^{[{\rm JC}'](S)}_{q{\bar q}}({\hat \tau}).
\end{eqnarray*}

The calculated results are given by 
\begin{eqnarray*}
I_{q{\bar q}(NS)}^{[{\rm JC}'](S)}&&(1,1-\eta_s)= I_{q{\bar q}(NS)}^{[{\rm JC}'](S1)}(1,1-\eta_s)  \nonumber \\
& & = C_F\Big[-4SP_-(\eta_s,0)-2\log^2\eta_s +3\log{1-\eta_s \over \eta_s} 
 -{7 \over 2}+{4 \over 3}\pi^2\Big] 
\end{eqnarray*}
for $0 < \eta_s \leq {\hat \tau}_M$ and  
\begin{eqnarray*}
I_{q{\bar q}(NS)}^{[{\rm JC}'](S)}&&(1,1-\eta_s)= I_{q{\bar q}(NS)}^{[{\rm JC}'](S1)}(1,1-{\hat \tau}_M)+I_{q{\bar q}(NS)}^{[{\rm JC}'](S2)}(1-{\hat \tau}_M,1-\eta_s)  
\end{eqnarray*}
with
\begin{eqnarray*}
I_{q{\bar q}(NS)}^{[{\rm JC}'](S2)}&&(1-{\hat \tau}_M,1-\eta_s)= 2C_F\Big[2SP_-\left({\eta_s-{\hat \tau}_M \over 1-{\hat \tau}_M},0\right)+2SP_+\left({\eta_s-{\hat \tau}_M \over {\hat \tau}_M},0\right)-SP_-(\eta_s,{\hat \tau}_M) \nonumber \\
& & -\log{1-\eta_s \over 1-{\hat \tau}_M}\left(\log{{\hat \tau}_M \over (1-{\hat \tau}_M)^2}+1-{\hat \tau}_M\right)+2(\eta_s-{\hat \tau}_M)+\eta_s\log{{\hat \tau}_M \over \eta_s}\Big] 
\end{eqnarray*}
for ${\hat \tau}_M < \eta_s \leq 2{\hat \tau}_M$.

 Here, we define 
\begin{eqnarray*}
SP_{\pm}(a,b) \equiv \int^a_bd{\hat \tau}{\log{\hat \tau} \over 1 \pm {\hat \tau}}.
\end{eqnarray*}

The contribution of the hard gluon radiation ($I=H$) integrated over the range $M^2 \leq -{\hat t} \leq {\hat s}(1-{\hat \tau})-M^2$ and ${\hat \tau}_0 \leq {\hat \tau} < 1-2{\hat \tau}_M$ is given by 
\begin{eqnarray*}
\sigma_{q{\bar q}}^{(H)} (1-2{\hat \tau}_M,{\hat \tau}_0)\simeq{{\bar \alpha}_s({\hat s},1-2{\hat \tau}_M,{\hat \tau}_0) \over 2\pi}{\hat \sigma}_0({\hat s},0)I_{q{\bar q}}^{(H)} (1-2{\hat \tau}_M,{\hat \tau}_0)
\end{eqnarray*}
with
\begin{eqnarray*}
I_{q{\bar q}}^{(H)}&&(1-2{\hat \tau}_M,{\hat \tau}_0)=2C_F\Big[2SP_+\left({1-{\hat \tau}_M-{\hat \tau}_0 \over {\hat \tau}_M},1\right)+SP_-\left({1-{\hat \tau}_M-{\hat \tau}_0 \over 1-{\hat \tau}_M},{{\hat \tau}_M \over 1-{\hat \tau}_M}\right) \\
& & +\log{1-2{\hat \tau}_M \over {\hat \tau}_0}\left(\log{1-{\hat \tau}_M \over { \hat \tau}_M}-1+2{\hat \tau}_M \right) 
 -\left(1-{\hat \tau}_M-{\hat \tau}_0\right)\log{1-{\hat \tau}_M-{\hat \tau}_0 \over {\hat \tau}_M}+2\left(1-2{\hat \tau}_M-{\hat \tau}_0\right)\Big],
\end{eqnarray*}
with an averaged coupling constant ${\bar \alpha}_s({\hat s},a,b)\equiv(\alpha_s(a{\hat s})+\alpha_s(b{\hat s}))/2$, where $a$ and $b$ are the upper and  lower limits of ${\hat \tau}$ integration, respectively.

\end{document}